\documentclass[aps,prd,showpacs,preprintnumbers,twocolumn,groupedaddress,nofootinbib]{revtex4-1}
\pdfoutput=1

\usepackage[]{amsmath}
\usepackage[]{amssymb}
\usepackage[]{mathrsfs}
\usepackage[]{eucal}
\usepackage[]{tensor}
\usepackage[]{amsthm}
\usepackage[]{bm}

\usepackage{tikz}

\usepackage[unicode]{hyperref}
\usepackage{xcolor}
\definecolor{pastgreen}{HTML}{669900}
\definecolor{pastblue}{HTML}{336699}
\definecolor{linkcol}{HTML}{663333}
\hypersetup{colorlinks,linkcolor={pastblue},citecolor={pastblue},urlcolor={pastblue}}

\newcommand{\pf}[1]{\mathbf{#1}}
\newcommand{\dd}{\partial}
\newcommand{\hdg}{\star} 
\newcommand{\df}{\mathrm{d}}
\newcommand{\w}{\wedge}
\newcommand{\veps}{\bm{\epsilon}}
\newcommand{\Lie}{\pounds}
\newcommand{\nab}[1]{\nabla_{\!#1}}

\newcommand{\qqd}{\ , \quad}
\newcommand{\bc}{\begin{center}}
\newcommand{\ec}{\end{center}}
\newcommand{\be}{\begin{equation}}
\newcommand{\ee}{\end{equation}}

\newcommand{\defeq}{\mathrel{\mathop:}=}

\newcommand{\MM}{\mathscr{M}}
\newcommand{\FF}{\mathcal{F}}
\newcommand{\GG}{\mathcal{G}}
\newcommand{\LL}{\mathscr{L}}
\newcommand{\scr}[1]{\mathrm{\scriptscriptstyle{#1}}}

\usepackage{dsfont}

\newcommand{\rr}{\mathds{R}}

\newcommand{\cl}[1]{\overline{#1}}

\theoremstyle{plain} \newtheorem{tm}{Theorem}[]
\newcommand{\btm}{\begin{tm}}
\newcommand{\etm}{\end{tm}}

\begin{document} 
\begin{flushleft}
\texttt{ZTF-EP-21-07}

\texttt{RBI-ThPhys-2021-41}
\end{flushleft}

\title{Nonlinear electromagnetic fields in strictly stationary spacetimes}

\author{A. Bokuli\'c}
\email{abokulic@phy.hr}
\affiliation{Department of Physics, Faculty of Science, University of Zagreb, 10000 Zagreb, Croatia}
\author{T. Juri\'c}
\email{tjuric@irb.hr}
\affiliation{Rudjer Bo\v skovi\'c Institute, Bijeni\v cka cesta 54, HR-10002 Zagreb, Croatia}
\author{I. Smoli\'c}
\email{ismolic@phy.hr}
\affiliation{Department of Physics, Faculty of Science, University of Zagreb, 10000 Zagreb, Croatia}

\begin{abstract}
We prove two theorems which imply that any stationary nonlinear electromagnetic field obeying a dominant energy condition in a strictly stationary, everywhere regular, asymptotically flat spacetime must be either trivial or a stealth field. The first theorem holds in static spacetimes and is independent of the gravitational part of the action, as long as the coupling of the electromagnetic field to the gravitational field is minimal. The second theorem assumes Einstein--Hilbert gravitational action and relies on the positive energy theorem, but does not assume that the spacetime metric is static. In addition, we discuss possible generalizations of these results, to theories with charged matter, as well as higher-dimensional nonlinear electromagnetic fields.
\end{abstract}

\maketitle

\section{Introduction} 

Interaction of the gravitational and the electromagnetic fields, governed by the gravitational-gauge field equations, is highly nonlinear. It is quite optimistic to hope that we might reach a complete classification of all solutions, even under the constraints of some regularity and boundary conditions. Indeed, a slightly less ambitious goal, understanding of time-independent solutions, is still a formidable task, but one worth taking as stationary solutions serve as models of the equilibrium field configurations. 

\smallskip

For example, an important class of \emph{stationary} black hole spacetimes is heavily narrowed by the series of black hole uniqueness and no-hair theorems \cite{HCC,Heusler}, distilled and polished over the past several decades. These solutions, however, are not \emph{strictly} stationary, as the Killing vector field corresponding to stationary isometry, timelike on some domain of the black hole exterior, may change its causal character in the black hole interior and ergoregions surrounding rotating black holes. Also, black holes may harbor a singularity, in which case they are not globally regular spacetimes.

\smallskip

This begs the question whether it is possible to have a strictly stationary, everywhere regular, asymptotically flat solution with a nonvanishing electromagnetic field. Such a spacetime would represent an instance of Wheeler's gravitational-electromagnetic geon \cite{Wheeler55}, at least up to a nontrivial question of stability. A negative answer in the case of Einstein--Maxwell theory is a canonical, well-known result, sometimes referred to as the absence of self-gravitating electromagnetic solitons \cite{HCC}. Setting aside a delicate historical question of primacy, the basic strategy of proofs can be traced back to the seminal work of Lichnerowicz \cite{Lichnerowicz}: construct a convenient non-negative quantity, whose integral over the spacetime domain in the problem is nonpositive, implying that this quantity has to be identically zero. This was masterfully utilized in the foundational uniqueness theorems obtained by Carter \cite{C73} (cf.~republished, corrected paper \cite{C73r}) and, more recently, by Heusler \cite{Heusler94,Heusler}. Several generalizations of the ``no-soliton theorem'' for the Einstein--Maxwell theory in the presence of various scalar fields was obtained by Shiromizu, Ohashi and Suzuki \cite{SOS12}, and Herdeiro and Oliveira \cite{HO19,HO20}.

\smallskip

One step further is to ask what happens in the theories where Maxwell's classical electrodynamics is replaced by its nonlinear modifications. Nonlinear electrodynamics (NLE) has its roots at the dawn of quantum field theory, back in the 1930s, sprouting over the following decades with innumerable NLE Lagrangians. Born--Infeld theory \cite{Born34,BI34} was constructed with the specific aim to cure the inconsistencies of Maxwell's electrodynamics associated with the infinite self-energy of the point charges and, remarkably, reappeared much later in low energy limits of the string theory \cite{FT85}. Another prominent NLE theory is defined by the Euler--Heisenberg one-loop QED correction to Maxwell's Lagrangian \cite{HE36,Dunne04}. The repository of proposed NLE theories has been growing ever since, with Lagrangian densities constructed from logarithmic \cite{Soleng95}, hyperbolic tangent \cite{ABG99}, power \cite{HM07,HM08}, exponential function \cite{Hendi13}, and so forth. Novel ModMax electrodynamics \cite{BLST20,Kosyakov20} is a one-parameter class of NLE theories which is both conformally invariant and invariant with respect to electromagnetic duality rotations \cite{GR95}. Nonlinearities in the electromagnetic interaction are being tested by the ATLAS Collaboration \cite{ATLAS17,EMY17,NAM17,NAM18} and new generations of the ultraintense lasers at the Extreme Light Infrastructure \cite{ELINP20}.

\smallskip

An intriguing feature of NLE theories is that they may admit a resolution of the black hole singularities, up to delicate constraints \cite{BMSS79,Bronnikov00,BH02,Bronnikov17}. An example of a regular black hole spacetime, originally proposed \emph{ad hoc} by Bardeen \cite{Bardeen68}, was later interpreted by Ay\'on-Beato and Garc\'ia \cite{ABG98,ABG00} as a solution of Einstein-NLE-Maxwell field equations for a particular NLE theory (broader analyses of spherically symmetric solutions may be found in \cite{deO94,AL06,DARG09,DARG10}; cf.~also \cite{TZ10,EP21}). Recently found, an even less trivial, regular black hole solution \cite{CM21} is based on a NLE theory with nonminimal coupling of the electromagnetic field to the gravitation. Nevertheless, these examples still leave the original question open: will a self-gravitating electromagnetic field settle in a nontrivial, regular configuration which is not a black hole? The first extension of the ``no-soliton'' theorem (referred to by the authors as the ``Lichnerowicz-type theorem'') for theories with NLE was given in \cite{CPX14} for the truncated Born--Infeld theory and the power-Maxwell theory. Our aim is to provide a much broader generalization of this result for NLE Lagrangians which are general smooth functions of both electromagnetic invariants, $F_{ab} F^{ab}$ and $F_{ab} \, {\hdg F}^{ab}$. Also, we shall present the first steps in the generalization of these results for theories with charged matter or theories in a different number of spacetime dimensions.

\smallskip

The paper is organized as follows. In Sec. II we briefly overview the fundamentals of gravitational theories with nonlinear electromagnetic fields. The main results of the paper, theorems \ref{tm1} and \ref{tm2}, are stated in Sec. III and their proofs are presented in Sec. IV. We discuss various generalizations of these theorems in Sec. V and review the remaining open questions in Sec. VI. Several basic identities from differential geometry are stated in the Appendix.

\smallskip

\emph{Conventions and notation}. The interior of a set $S$ is denoted by $S^\circ$, the boundary of $S$ by $\dd S$ and the closure of $S$ by $\cl{S}$. The difference between sets $A$ and $B$ is denoted by $A - B$. We shall use the ``mostly plus'' metric signature and the natural system of units with $G = c = 4\pi\epsilon_0 = 1$. Differential forms are denoted by bolded indexless letters, an abstract index notation or a combination of both. The volume 4-form is denoted by $\veps$. The contraction of a symmetric tensor $S_{ab}$ with vector $X^a$ is a 1-form denoted by $\pf{S}(X)$. Following reference \cite{HCC}, we write $f = O(r^{-k})$ when $f$ is of order $O(r^{-k})$ as $r \to \infty$ and $f = O_\infty(r^{-k})$ when $\dd_{i_1} \dots \dd_{i_\ell} f = O(r^{-k-\ell})$ for an arbitrary set of coordinate indices $\{i_1,\dots,i_\ell\}$.

\section{Brief overview of NLE} 

Let us, before stating the central theorems of the paper, briefly introduce the nonlinear electrodynamics. The ubiquitous elements are two electromagnetic invariants,
\be
\FF \defeq F_{ab} F^{ab} \quad \textrm{and} \quad \GG \defeq F_{ab} \, {\hdg F}^{ab} \ .
\ee
We follow the nomenclature from \cite{BJS21} by sorting NLE theories into the $\FF$-class, with a Lagrangian density $\LL$ depending only on invariant $\FF$, and the $\FF\GG$-class, with Lagrangian density $\LL$ depending on both invariants. In this paper the main focus is on the broader, $\FF\GG$-class of NLE theories, with the NLE Lagrangian density $\LL$ which is a $C^2$ function on some neighbourhood of the origin of the $\FF$-$\GG$ plane. We can always choose the Lagrangian density, by adding an appropriate constant, such that $\LL(0,0) = 0$. Partial derivatives of the Lagrangian density $\LL$ are denoted by abbreviations such as $\LL_\FF \defeq \dd_\FF\LL$, $\LL_\GG \defeq \dd_\GG\LL$, $\LL_{\FF\GG} \defeq \dd_\GG\dd_\FF \LL$, and so on. We say that a NLE Lagrangian density $\LL$ obeys Maxwell's weak field limit if $\LL_\FF(0,0) = -1/4$ and $\LL_\GG(0,0) = 0$. 

\smallskip

The Lagrangian 4-form, defined with some (diffeomorphism covariant) gravitational Lagrangian density $\LL^\scr{(g)}$, is
\be\label{eq:Lform}
\pf{L} = \frac{1}{16\pi} \, \big( \LL^\scr{(g)} + 4\LL \big) \, \veps \ .
\ee
The corresponding gravitational field equation is of the form
\be\label{eq:Einst}
E_{ab} = 8\pi T_{ab} \ ,
\ee
where the gravitational tensor $E_{ab}$ is divergence-free, $\nabla^a E_{ab} = 0$, and the NLE energy-momentum tensor may be conveniently written as
\be\label{TNLE}
T_{ab} = -4\LL_\FF T_{ab}^{\scr{(Max)}} + \frac{1}{4}\,T g_{ab} 
\ee
with Maxwell's electromagnetic energy-momentum tensor
\be
T_{ab}^{\scr{(Max)}} \defeq \frac{1}{4\pi} \left( F_{ac} \tensor{F}{_b^c} - \frac{1}{4} \, g_{ab} \FF \right) 
\ee
and the trace
\be
T \defeq g^{ab} T_{ab} = \frac{1}{\pi} \left( \LL - \LL_\FF \FF - \LL_\GG \GG \right) \ .
\ee
For an Einstein--Hilbert Lagrangian density $\LL^\scr{(g)} = R$ we have $E_{ab} = G_{ab}$, the Einstein tensor.

\smallskip

Using an auxiliary 2-form
\be
\pf{Z} \defeq -4 \left( \LL_\FF \pf{F} + \LL_\GG \, {\hdg\pf{F}} \right) \ ,
\ee
the NLE Maxwell's equations may be written in the form
\be\label{eq:NLEMax}
\df \pf{F} = 0 \qqd \df {\hdg\pf{Z}} = 0 \ .
\ee
We shall invoke two energy conditions, the null energy condition (NEC) and the dominant energy condition (DEC). It can be shown \cite{Plebanski70,BJS21} that the NEC holds if and only if $\LL_\FF \le 0$, while the DEC holds if and only if $\LL_\FF \le 0$ and $T \le 0$.

\section{Two theorems} 

We shall present two ``no-soliton'' theorems, each of which has its strengths and limitations. Both theorems assume that the spacetime is strictly stationary, so that we do not consider spacetimes with either black hole or cosmological horizons. The first result rests upon a stronger assumption, that spacetime is static, which admits a simpler proof that does not depend on details of the gravitational Lagrangian of the theory, as long as the coupling of the electromagnetic field to gravitation is minimal. The second result does not rely on this assumption, so that it may be applied to ``rotating'' solutions. However, this comes at a price: proof rests upon a highly nontrivial, celebrated positive energy theorem \cite{SY79a,SY79b,SY81a,SY81b,Witten81,PT82,Dain13} and, correspondingly, works only in those gravitational theories for which this theorem has been proven.

\smallskip

Let us, before the statement of the main results, list the technical assumptions necessary for the theorems.

\begin{itemize}
\item[(1)] The spacetime consists of a four-dimensional smooth, simply connected manifold $\MM$, with a smooth Lorentzian metric $g_{ab}$ and a smooth electromagnetic 2-form $F_{ab}$, which are solutions of the gravitational-NLE field equations (\ref{eq:Einst}) and (\ref{eq:NLEMax}), with the NLE Lagrangian density $\LL$ obeying Maxwell's weak field limit.

\item[(2)] The spacetime admits a strictly timelike Killing vector field $k^a$ (namely, $k^a k_a < 0$ on the whole $\MM$) and the electromagnetic field inherits the symmetry $\Lie_k F_{ab} = 0$ \cite{Tod06,CDPS16,BGS17}. 

\item[(3)] Through each point $p \in \MM$ passes at least one complete oriented spacelike hypersurface $\Sigma$ with induced metric $h_{ij}$ and the associated second fundamental form (extrinsic curvature) $K_{ij}$, Euclidean at infinity\footnote{We say that a smooth $n$-manifold $S$ is \emph{Euclidean at infinity} if there is a compact set $\mathcal{C} \subseteq S$, such that $S - \mathcal{C}$ is a disjoint union of a finite number of sets (``ends''), each of which is diffeomorphic to the complement of a contractible compact set in $\rr^n$.} and asymptotically flat in the sense \cite{PT82,Dain13,HCC} that on each of its ``ends'' the following fall-off conditions, written in Cartesian coordinates, are met: $1 + k^\alpha k_\alpha = O_\infty(r^{-1})$, $k^\alpha g_{\alpha i} = O_\infty(r^{-1})$, $\gamma_{ij} = O_\infty(r^{-1})$ and $K_{ij} = O_\infty(r^{-2})$, while the electromagnetic 2-form $F_{ab}$ satisfies $k^\alpha F_{\alpha i} = O_\infty(r^{-2})$ and $k^\alpha {\hdg F}_{\alpha i} = O_\infty(r^{-2})$, and the associated potentials (defined below) are of order $O_\infty(r^{-1})$.
\end{itemize}

\smallskip

\noindent
We shall refer to assumptions (1)--(3) as the \emph{basic assumptions}. It is quite possible that some of the assumptions above may be slightly relaxed without any significant effect on the further conclusions, but we shall not pursue such nuances here. Furthermore, we introduce the following notion.

\smallskip

\noindent
\textbf{Definition}. We say that an electromagnetic field is \emph{stealth} at a point $p \in \MM$ if at that point the corresponding energy-momentum tensor $T_{ab}$ is zero, but the electromagnetic 2-form $F_{ab}$ is nonzero.

\smallskip

In other words, stealth fields do not affect the spacetime metric, as their contribution to the energy-momentum tensor is vanishing. Such configurations are absent in Maxwell's classical electrodynamics, but appear in NLE theories if and only if $F_{ab} \ne 0$, $\LL_\FF = 0$ and $T = 0$ hold at a given point \cite{ISm18}. Note that in this paper, for clarity, we keep the trivial fields, $F_{ab} = 0$, apart from the stealth fields. A prominent class of stealth solution examples \cite{ISm18} may be found among the null electromagnetic fields in power-Maxwell theory \cite{HM07,HM08}, which also belong to a family of so-called universal electromagnetic fields \cite{OP15,OP18,HOP18}.

\smallskip

The two central results of the paper are as follows.

\btm\label{tm1}
Suppose that a spacetime with an electromagnetic field satisfies basic assumptions, with the electromagnetic energy-momentum tensor obeying the null energy condition, and where the Killing vector field $k^a$ is hypersurface orthogonal. Then the electromagnetic field is at each point of the spacetime either trivial, $F_{ab} = 0$, or stealth.
\etm

\btm\label{tm2}
Suppose that a spacetime with the electromagnetic field satisfies basic assumptions and the gravitational part of the action is the Einstein--Hilbert's with the electromagnetic energy-momentum tensor obeying the dominant energy condition. Then the spacetime is isometric to the Minkowski spacetime $(\rr^4,\eta_{ab})$ and the electromagnetic field is at each point of the spacetime either trivial, $F_{ab} = 0$, or stealth.
\etm

\smallskip

We stress that theorem \ref{tm1} relies on a weaker, null energy condition.

\section{Proofs of theorems} 

In both theorems we are looking at a spacetime admitting a strictly timelike Killing vector field $k^a$. It is convenient to introduce the function $V \defeq -k_a k^a > 0$ and the associated twist 1-form
\be
\bm{\omega} \defeq -{\hdg (\pf{k} \w \df\pf{k})} \ .
\ee
One should, however, beware of variations in the definition of the twist 1-form throughout the literature (e.g.~Heusler \cite{Heusler} introduces a twist 1-form $\widetilde{\bm{\omega}}$, such that $\bm{\omega} = -2\widetilde{\bm{\omega}}$). Our choice is mainly motivated by the fact that, in the abstract index notation, it corresponds simply to $\omega_a = \tensor{\epsilon}{_a^b^c^d} k_b \nab{c} k_d$, without additional factors.

\smallskip 

The vector field $k^a$ allows us to define two electric 1-forms,
\be
\pf{E} \defeq -i_k \pf{F} \qqd \pf{D} \defeq -i_k \pf{Z} \ ,
\ee
and two magnetic 1-forms,
\be
\pf{B} \defeq i_k {\hdg \pf{F}} \qqd \pf{H} \defeq i_k {\hdg \pf{Z}} \ .
\ee
As a consequence of the symmetry inheritance and generalized Maxwell's equations we know that $\pf{E}$ and $\pf{H}$ are closed forms,
\begin{align}
\df\pf{E} & = -\df i_k \pf{F} = (-\Lie_k + i_k \df) \pf{F} = 0 \ ,\\
\df\pf{H} & = \df i_k {\hdg\pf{Z}} = (\Lie_k - i_k \df) {\hdg\pf{Z}} = 0 \ .
\end{align}
As manifold $\MM$ is, by assumption, simply connected, we can globally define scalar potentials, electric $\Phi$ and magnetic $\Psi$, such that $\pf{E} = -\df\Phi$ and $\pf{H} = -\df\Psi$. Also, directly from the definition we know that $\Lie_k \Phi = -i_k \pf{E} = 0$ and $\Lie_k \Psi = -i_k \pf{H} = 0$.

\smallskip

The backbone of the proofs are divergence identities which have to be carefully chosen. First, using
\be
\df\left( \frac{\pf{k}}{V} \right) = \frac{1}{V^2} \left( V \df\pf{k} - \df V \w \pf{k} \right) = \frac{1}{V^2} \, {\hdg(\bm{\omega} \w \pf{k})}
\ee
and
\be
-{\hdg \, i_k \pf{Z}} = {\hdg \, i_k {{\hdg\hdg}\pf{Z}}} = \pf{k} \w {\hdg\pf{Z}} 
\ee
we have
\begin{align}
\nabla^a \left( \frac{D_a}{V} \right) & = -{\hdg\df\hdg} \left( -\frac{1}{V}\,i_k \pf{Z} \right) \nonumber \\ 
 & = -{\hdg\df\left( \frac{1}{V} \, \pf{k} \w {\hdg\pf{Z}} \right)} \nonumber\\
 & = -{\hdg \left( \frac{1}{V^2} \, {\hdg(\bm{\omega} \w \pf{k}) \w {\hdg\pf{Z}}} \right)} \nonumber\\
 & = \frac{1}{2V^2} \, (\bm{\omega} \w \pf{k})_{ab} \, {\hdg\pf{Z}}^{ab} \ .
\end{align}
Therefore,
\be
\nabla^a \left( \frac{D_a}{V} \right) = -\frac{\omega_a H^a}{V^2}
\ee
and, analogously,
\be
\nabla^a \left( \frac{B_a}{V} \right) = \frac{\omega_a E^a}{V^2} \ .
\ee

\medskip

\emph{Proof of theorem \ref{tm1}}. Let us introduce an auxiliary open set
\be
O \defeq \left\{ x \in \MM \mid \LL_\FF(x) \ne 0 \right\} \ .
\ee
In other words, due to the assumed NEC, $\LL_\FF(x) < 0$ for all $x \in O$ and $\LL_\FF(y) = 0$ for all $y \in \MM - O$. As the electromagnetic field decays along each ``end'' and the Lagrangian density obeys the Maxwellian weak field limit, we know that $O$ is nonempty. 

\smallskip

At each point of the complement $\MM - O$, the gravitational field equation is reduced to $E_{ab} = 2\pi T g_{ab}$. Thus, the divergence $\nabla^a E_{ab} = 0$ implies that the trace $T$ is constant on each connected component of the interior $(\MM-O)^\circ$. Furthermore, by the assumption of the theorem, $\bm{\omega} = 0$, so that $k^a$ is a hypersurface orthogonal vector field. Let $\Sigma$ be an arbitrary spacelike hypersurface from the basic assumption (3). Inserting, respectfully, $\bm{\alpha} = \pf{D}/V$ and $\bm{\alpha} = \pf{B}/V$ in Eq.~(\ref{eq:intSigma}), both of which satisfy $\Lie_k \bm{\alpha} = 0$, we get
\be\label{eq:kDkB}
\int_{\dd\Sigma} \frac{1}{V} \, {\hdg(\pf{k} \w \pf{D})} = 0 \qqd \int_{\dd\Sigma} \frac{1}{V} \, {\hdg(\pf{k} \w \pf{B})} = 0 \ .
\ee
Formally, the integral over $\dd\Sigma$ may denote the limit for the integral over the ``sphere at infinity''. Furthermore, for $\bm{\alpha} = \Phi\pf{D}/V$ and $\bm{\alpha} = \Psi\pf{B}/V$ we get, respectfully,
\be
\nabla^a \left( \frac{\Phi}{V} \, D_a \right) = \frac{4}{V} \left( \LL_\FF E_a E^a - \LL_\GG E_a B^a \right)
\ee
and
\be
\nabla^a \left( \frac{\Psi}{V} \, B_a \right) = \frac{4}{V} \left( \LL_\FF B_a B^a + \LL_\GG E_a B^a \right) \ .
\ee
The sum of these two equations,
\be
\nabla^a \left( \frac{\Phi}{V} \, D_a + \frac{\Psi}{V} \, B_a \right) = \frac{4}{V} \left( \LL_\FF E_a E^a + \LL_\FF B_a B^a \right)
\ee
integrated over $\Sigma$, with the help of (\ref{eq:kDkB}) and the fall-off conditions on potentials $\Phi$ and $\Psi$, leads to
\be
\int_\Sigma \frac{\LL_\FF}{V} \left( E_a E^a + B_a B^a \right) \hat{\veps} = 0 \ ,
\ee
with the induced volume 3-form $\hat{\veps}$. Now, as $\pf{k}$ is strictly timelike, $V > 0$, neither $E^a$ nor $B^a$ can be causal (as $k_a E^a = 0$ and $k_a B^a = 0$), and the integrand above is nonpositive on $O \cap \Sigma$ and zero on $(\MM - O) \cap \Sigma$. As the total integral is zero, it follows that $E^a = 0 = B^a$ and, as $\LL(0,0) = 0$, consequently $T = 0$ on $O \cap \Sigma$. By continuity this implies that $T = 0$ on $\cl{O} \cap \Sigma$, and thus $T = 0$ on the whole $\Sigma$. In conclusion, on each point of the set $O \cap \Sigma$ we have $F_{ab} = 0$, while on each point of the set $(\MM - O) \cap \Sigma$ the electromagnetic field $F_{ab}$ is either zero or stealth.

\smallskip

We stress that the argument works irrespectively of the gravitational part of the equations of motion, as long as the coupling is minimal and tensor $E_{ab}$ is divergence-free. If $\LL_\FF = 0$ we can find simple counterexamples, such as the stealth field on static background \cite{ISm18}.

\medskip

\emph{Proof of theorem \ref{tm2}}. Here we turn to the Einstein--Hilbert case, $E_{ab} = G_{ab}$. The exterior derivative of the twist 1-form, with the help of the Killing lemma $\df{\hdg\df\pf{k}} = 2\,{\hdg\pf{R}(k)}$, may be written as
\begin{align}
\df\bm{\omega} & = -2{\hdg(\pf{k} \w \pf{R}(k))} = \nonumber\\
 & = 64\pi\LL_\FF \, {\hdg(\pf{k} \w \pf{T}^\scr{(Max)}(k))} = \nonumber\\
 & = 4 \pf{E} \w \pf{H} \ .
\end{align}
Using electromagnetic scalar potentials,
\be
\df\bm{\omega} = -4(\df\Phi \w \pf{H}) = -4(\pf{E} \w \df\Psi)
\ee
we see that both $\bm{\omega} + 4\Phi\pf{H}$ and $\bm{\omega} - 4\Psi\pf{E}$ are closed 1-forms. Thus, as $\MM$ is, by assumption, simply connected, we can globally define the new scalar potentials $U_E$ and $U_H$ such that
\begin{align}
\bm{\omega} & = -4\Phi\pf{H} + \df U_H \\
 & = 4\Psi\pf{E} + \df U_E \ .
\end{align}
As with the electromagnetic potentials $\Phi$ and $\Psi$, we see directly from the definition that $\Lie_k U_E = 0$ and $\Lie_k U_H = 0$. Note that
\begin{align}
\omega_a \omega^a & = -4\Phi \omega_a H^a + \omega^a \nab{a} U_H \\
 & = 4\Psi \omega_a E^a + \omega^a \nab{a} U_E \ .
\end{align}
Now, using a basic formula
\be
\nabla^a \left( \frac{\omega_a}{V^2} \right) = 0
\ee
and the relations above, we have four divergence identities,
\begin{align}
\nabla^a \left( U_E \, \frac{\omega_a}{V^2} \right) & = \frac{\omega^a \nab{a} U_E}{V^2} \ , \\
\nabla^a \left( U_H \, \frac{\omega_a}{V^2} \right) & = \frac{\omega^a \nab{a} U_H}{V^2} \ , \\
\nabla^a \left( \frac{\Phi}{V}\,D_a \right) & = \frac{4}{V} \, (\LL_\FF E_a E^a - \LL_\GG E_a B^a) - \Phi\,\frac{\omega_a H^a}{V^2} \ , \\
\nabla^a \left( \frac{\Psi}{V}\,B_a \right) & = \frac{4}{V} \, (\LL_\FF B_a B^a + \LL_\GG E_a B^a) + \Psi\,\frac{\omega_a E^a}{V^2} \ .
\end{align}
A brief inspection reveals that an auxiliary 1-form
\be
\pf{W} \defeq \frac{U_E + U_H}{V^2} \, \bm{\omega} + \frac{4}{V} \, (\Phi\pf{D} + \Psi\pf{B})
\ee
has a rather simple covariant divergence,
\be
\nab{a} W^a = \frac{16}{V} \, \LL_\FF \left( E_a E^a + B_a B^a \right) + 2\,\frac{\omega_a \omega^a}{V^2} \ .
\ee
Taking into account Einstein's field equation,
\be
R_{ab} = 8\pi \left( T_{ab} - \frac{1}{2} \, T g_{ab} \right)
\ee
and
\be
8\pi T_{ab}^{\scr{(Max)}} k^a k^b = E_a E^a + B_a B^a \ ,
\ee
we get
\be\label{eq:Rkk}
\frac{4}{V} \, R_{ab} k^a k^b - \frac{2}{V^2} \, \omega_a \omega^a = -\nab{a} W^a + 8\pi T \ .
\ee
Now we turn to Heusler's mass formula \cite{Heusler94}. If we contract
\be
{\hdg\df\bm{\omega}} = 2\,\pf{k} \w \pf{R}(k)
\ee
with $k^a$ and take the Hodge dual, we obtain
\be
-{\hdg\pf{R}(k)} = \frac{R_{ab}k^a k^b}{V} \, {\hdg\pf{k}} + \frac{1}{2V} \, \pf{k} \w \df\bm{\omega} \ .
\ee
Also, using 
\be
-\df\left( \frac{1}{V} \, \pf{k} \w \bm{\omega} \right) = \frac{\omega_a \omega^a}{V^2} \, {\hdg\pf{k}} + \frac{1}{V} \, \pf{k} \w \df\bm{\omega}
\ee
we have
\be
-{\hdg\pf{R}(k)} = \left(\frac{R_{ab}k^a k^b}{V} - \frac{\omega_a \omega^a}{2V^2} \right) {\hdg\pf{k}} - \df\left( \frac{1}{2V} \, \pf{k} \w \bm{\omega} \right) \ .
\ee
Thus, relying on the fall-off properties of the twist 1-form $\bm{\omega}$ inferred from the basic assumptions, Komar's mass 
\be
M = -\frac{1}{4\pi} \int_\Sigma {\hdg\pf{R}(k)}
\ee
may be written in the form
\be
M = \frac{1}{4\pi} \, \int_\Sigma \left( \frac{R_{ab}k^a k^b}{V} - \frac{\omega_a \omega^a}{2V^2} \right) {\hdg \pf{k}} \ ,
\ee
which, in combination with (\ref{eq:Rkk}), becomes
\be
M = -\frac{1}{16\pi} \int_\Sigma \nab{a} W^a \, {\hdg \pf{k}} + \frac{1}{2} \int_\Sigma T \, {\hdg \pf{k}} \ .
\ee
We note in passing that the formula obtained here is consistent with the generalized Smarr formula \cite{GS17,ZG18,BJS21}. The $W$-term vanishes at infinity, while the $T$-term is nonpositive, given that the DEC holds. Finally, the positive energy theorem implies that $M \ge 0$ and $M = 0$ if and only if the spacetime is Minkowski. As we have proven that $M \le 0$, it follows that $M = 0$. Therefore $T_{ab} = 0$, implying that any nontrivial electromagnetic field must be zero or stealth.

\section{Further generalizations} 

We now turn to possible generalizations of the main results, which we shall sort into two directions.

\smallskip

\emph{Theories with charged matter}. As the simplest model of matter we may choose a complex scalar field $\phi$, with the total Lagrangian
\be
\LL^{\mathrm{(tot)}} = \LL(\FF,\GG) + (\mathcal{D}_a \phi)^*(\mathcal{D}^a \phi) - \mathscr{U}(\phi^*\phi) \ ,
\ee
constructed with the covariant gauge derivative $\mathcal{D}_a = \nab{a} + iq A_a$, the gauge 1-form $\pf{A}$ which defines the electromagnetic 2-form $\pf{F} = \df\pf{A}$, and the scalar (self-interaction) potential $\mathscr{U}$. In this theory generalized Maxwell's equations have the form
\be
\df{\hdg\pf{Z}} = 4\pi \, {\hdg\pf{J}}
\ee
with the current 1-form
\be
J_a = \frac{iq}{4\pi} \, \big( \phi^* \mathcal{D}_a \phi - \phi (\mathcal{D}_a \phi)^* \big) \ .
\ee
The electric 1-form $\pf{E}$ is again closed and we have the associated electric scalar potential $\Phi$. Let us, for simplicity, focus on the strictly static case, $\bm{\omega} = 0$. Then,
\be
\nabla^a \left( \frac{1}{V}\,D_a \right) = -\frac{4\pi}{V} \, k^a J_a \qqd \nabla^a \left( \frac{1}{V}\,B_a \right) = 0
\ee
and
\be\label{eq:divPsiVDJ}
\nabla^a \left( \frac{\Phi}{V}\,D_a \right) = \frac{4}{V} \, (\LL_\FF E_a E^a - \LL_\GG E_a B^a) - \frac{4\pi\Phi}{V} \, k^a J_a \ .
\ee
The first technical obstacle is a treatment of the term $\Phi k^a J_a$ which, without additional assumptions, is in general neither positive nor negative definite. Setting aside spacetimes with symmetry noninheriting scalar fields \cite{ISm15,ISm17,FS21}, let us for simplicity assume that $\Lie_k \phi = 0$. Also, taking into account remarks from \cite{BJS21,CDPS16}, we assume that the gauge choice is made such that $\Lie_k\pf{A} = 0$. Now, as
\be
\df(\Phi + i_k\pf{A}) = -\pf{E} + (\Lie_k - i_k\df)\pf{A} = 0 \ ,
\ee
given that both $\Phi$ and $k^a A_a$ vanish at infinity, we may set $\Phi = -k^a A_a$. This leads us to the simplification
\be
\Phi k^a J_a = \frac{(q\Phi)^2}{2\pi} \, {\phi^* \phi} \ge 0 \ .
\ee
Furthermore, the magnetic 1-form $\pf{H}$ is no longer necessarily a closed form as
\be
\df\pf{H} = 4\pi {\hdg(\pf{k} \w \pf{J})} \ .
\ee
This is a familiar obstacle to the introduction of magnetic scalar potential on domains which contain nonvanishing electric currents. There are several subcases in which we can proceed with a similar strategy of proof as above:
\begin{itemize}
\item[(a)] if $\pf{k} \w \pf{J} = 0$, which allows us to introduce the magnetic scalar potential $\Psi$ and deduce
\be
\nabla^a \left( \frac{\Psi}{V}\,B_a \right) = \frac{4}{V} \, (\LL_\FF B_a B^a + \LL_\GG E_a B^a) \ ;
\ee

\item[(b)] if we have a strictly electric system, in the sense that $\pf{B} = 0$, so that Eq.~(\ref{eq:divPsiVDJ}) may suffice for the proof.
\end{itemize}

Given that any of the two conditions above, (a) or (b), are met, repetition of the argument from the previous section leads to the conclusion that the electromagnetic field is trivial on the set $O$ where $\LL_\FF \ne 0$. We note in passing that Eq.~(\ref{eq:divPsiVDJ}) may be also used if we have an $\FF$-class theory, but only to deduce that $\pf{E} = 0$ on the set $O$, without any control of the magnetic field $\pf{B}$ (unless, again, we invoke condition (a) or (b)). 

\smallskip

Furthermore, the divergence of the gravitational field equation on the interior $(\MM - O)^\circ$ leads to $\nab{a} T = 4J^b F_{ba}$. If, in addition, we assume either (a) or (b) from above, the decomposition $V\pf{F} = \pf{k} \w \pf{E} + {\hdg(\pf{k} \w \pf{B})}$ and $k^a \nab{a} T = 0$ allow us to deduce $V \nab{a} T = 4(k^b J_b) E_a$. In particular, $\pf{B} = 0$ on the set $(\MM - O)^\circ$ also implies $\pf{D} = 0$ and, via divergence identities, $k^a J_a = 0$, leading to the conclusion that the trace $T$ is constant on each connected component of the domain $(\MM - O)^\circ$ (it is not clear if this necessarily holds in the (a) subcase).

\smallskip

In conclusion, at least under some additional assumptions, the initial problem can be reduced to the question of existence of self-gravitating scalar solitons \cite{Heusler,HR15}. It is important to stress that spacetimes with charged boson stars evade the partial no-go result from above due to the symmetry noninheriting scalar field. Namely, a typical ansatz for such solutions features a scalar field of the form $\phi(t,r) = f(r)e^{i\omega t}$, so that $\Lie_k \phi = i\omega \phi$ and the definiteness of the term $\Phi k^a J_a$ is in general lost.

\smallskip

\emph{Higher-dimensional theories}. Let us look at spacetimes of dimension $m \ge 5$. As the $\GG$ invariant is a scalar only in four spacetime dimensions, where $\pf{F}$ and its Hodge dual ${\hdg\pf{F}}$ are both 2-forms, here we treat only $\FF$-class theories. We may define a twist $(m-3)$-form as $\bm{\omega} \defeq (-1)^{m+1} {\hdg(\pf{k} \w \df\pf{k})}$, in order to preserve its form in abstract indices,
\be
\omega_{a_1 \dots a_{m-3}} = \tensor{\epsilon}{_{a_1}_{\dots}_{a_{m-3}}^b^c^d} k_b \nab{c} k_d \ ,
\ee
and the 1-form $\pf{D} \defeq -i_k \pf{Z}$ as above. Taking into account that
\be
i_k \pf{Z} = (-1)^{m+1} {\hdg(\pf{k} \w {\hdg\pf{Z}})}
\ee
we have
\be
\nabla^a \left( \frac{D_a}{V} \right) = \frac{1}{(m-2)!V^2} \, (\bm{\omega} \w \pf{k})_{a_1 \dots a_{m-2}} \, {\hdg\pf{Z}}^{a_1 \dots a_{m-2}} \ .
\ee
The basic divergence identity for a strictly static spacetime, with $\bm{\omega} = 0$, is
\be\label{eq:divPhiD}
\nabla^a \left( \frac{\Phi}{V} \, D_a \right) = \frac{4}{V} \, \LL_\FF \, E_a E^a \ .
\ee
Assuming the appropriate fall-off conditions, namely $\Phi = O(r^{-(m-3)})$ and $\pf{D} = O(r^{-(m-2)})$ (cf. also \cite{Ortaggio14}), we may repeat the previous argument to conclude that $\pf{E} = 0$ on the set $O$. A further technical obstacle is that in an $m$-dimensional spacetime magnetic fields $\pf{B} = i_k {\hdg\pf{F}}$ and $\pf{H} = i_k {\hdg\pf{Z}}$ are $(m-3)$-forms, so that we have less control on the sign of their squares, such as $B_{a_1 \dots a_{m-3}} B^{a_1 \dots a_{m-3}}$.

\section{Discussion} 

We have proved that, up to exotic stealth solutions, NLE theories on simply connected four-dimensional spacetimes do not admit globally regular, stationary solitonic solutions. These results admit only partial generalizations in the presence of the charged matter and in higher-dimensional theories. Limitations in both directions do not come as a surprise, due to known solutions with charged bosonic stars and an increased number of electromagnetic degrees of freedom with the number of spacetime dimensions.

\smallskip

The remaining open questions may be grouped as follows:

\begin{enumerate}
\item Are conclusions altered if the spacetime manifold $\MM$ is not simply connected?

\item How do we treat a NLE theory which does not obey the Maxwellian weak field limit?

\item Can constraints on four-dimensional theories with complex scalar fields and higher-dimensional theories be strengthened?
\end{enumerate}

Simple connectedness of the manifold $\MM$ was invoked in order to guarantee the existence of scalar potentials $\Phi$, $\Psi$, $U_E$ and $U_H$. This assumption is not necessary, but on a nonsimply connected manifold one needs to either (a) impose some boundary conditions which imply the existence of scalar potentials, or (b) construct divergence identities which do not involve scalar potentials and which allow one to prove theorems analogous to the ones treated in this paper.

\smallskip

The Maxwellian weak field limit or a, slightly weaker, mere assumption that partial derivatives $\LL_\FF$ and $\LL_\GG$ are well-defined and finite at the origin of the $\FF$-$\GG$ plane plays a role in the elimination of the boundary terms at asymptotic ends. However, there are NLE theories, such as the power-Maxwell theory \cite{HM07,HM08} (for powers less than $1$) and the ModMax theory \cite{GR95,BLST20,Kosyakov20,FAGMLM20}, which do not behave well in this sense. Here one must first find a proper way to incorporate some notion of asymptotic flatness, with the appropriate fall-off conditions for the fields.

\smallskip

We feel that conditions under which a NLE theory with a complex scalar field (or some other form of the charged matter) does not admit strictly stationary solitonic solutions should be mapped more carefully. Also, it is not clear how to sensibly choose additional assumptions which might lead to generalizations in higher-dimensional theories. Finally, we remark that in $(1+2)$-dimensional spacetimes one might again rely on the divergence relation (\ref{eq:divPhiD}), as well as the lower-dimensional positive energy theorem \cite{Wong12}, but the problem is that the natural fall-off condition for the scalar potential $\Phi = O(\ln r)$ does not seem to be sufficient to get rid of the boundary terms.

\smallskip

\appendix

\section{Menagerie of identities}

Let us first, for generality, assume that $(\MM,g_{ab})$ is a smooth $m$-dimensional Lorentzian manifold. The Hodge dual of a $p$-form $\bm{\alpha}$, defined as 
\be
(\hdg\alpha)_{a_{p+1} \dots a_m} \defeq \frac{1}{p!}\,\alpha_{a_1 \dots a_p} \tensor{\epsilon}{^{a_1}^{\dots}^{a_p}_{a_{p+1}}_{\dots}_{a_m}} \ ,
\ee
twice applied produces a sign according to
\be
{\hdg{\hdg\bm{\alpha}}} = (-1)^{p(m-p) + 1} \bm{\alpha} \ .
\ee 
A simple useful rule, so-called ``flipping over the Hodge'', reads
\be
i_X {\hdg\bm{\alpha}} = {\hdg(\bm{\alpha} \w \pf{X})} \ ,
\ee
where $\pf{X}$ is the associated 1-form, $X_a = g_{ab} X^b$.

\smallskip

We shall introduce an auxiliary coderivative operator, acting on $p$-form $\bm{\alpha}$ as
\be
\delta \bm{\alpha} \defeq (-1)^{m(p+1)+1} \, {\hdg\df{\hdg\bm{\alpha}}} \ ,
\ee
which in the abstract index notation is simply
\be
\delta \alpha_{a_1 \dots a_{p-1}} = \nabla^b \alpha_{b a_1 \dots a_{p-1}} \ .
\ee 
Note that for even $m$ we have $\delta \bm{\alpha} = -{\hdg\df{\hdg\bm{\alpha}}}$. If $K^a$ is a Killing vector field and $K_a = g_{ab} K^b$ the associated 1-form, then the following identity holds:
\be\label{eq:Liealpha}
\Lie_K \bm{\alpha} = \delta(\pf{K} \w \bm{\alpha}) + \pf{K} \w \delta\bm{\alpha} \ .
\ee
We are mostly interested in the case where $\bm{\alpha}$ is a 1-form such that $\Lie_K \bm{\alpha} = 0$. Whence, Eq.~(\ref{eq:Liealpha}) integrated over a smooth hypersurface $\Sigma$, with application of the generalized Stokes' theorem, leads to
\be\label{eq:intSigma}
\int_\Sigma (\delta\bm{\alpha}) \, {\hdg \pf{K}} = \int_{\dd\Sigma} {\hdg (\pf{K} \w \bm{\alpha})} \ ,
\ee
where we have, for simplicity, suppressed the pullback symbol.

\begin{acknowledgments}
The research was supported by the Croatian Science Foundation Project No.~IP-2020-02-9614.
\end{acknowledgments}

\bibliography{nogonle}

\end{document}